\documentclass[preprint,aps,12pt,showpacs,nofootinbib,tightenlines]{revtex4}
\usepackage{amsmath}
\usepackage{amssymb}
\usepackage{CJK}
\usepackage{epsfig}
\usepackage{graphicx}
\usepackage{subfigure}
\begin{document}
\begin{CJK*}{GBK}{song}
\def\pslash{\rlap{\hspace{0.02cm}/}{p}}
\def\eslash{\rlap{\hspace{0.02cm}/}{e}}
\title {Top quark pair production via (un)polarized photon collisions in the littlest Higgs model with T-parity at the
ILC}
\author{Bingfang Yang$^1$$^,$$^2$}\email{yangbingfang@gmail.com}
\affiliation{  $^1$College of Physics $\&$ Information Engineering,
Henan Normal University, Xinxiang 453007, China\\ $^2$ Basic
Teaching Department, Jiaozuo University, Jiaozuo 454000, China
   \vspace*{1.5cm}  }

\begin{abstract}
We study the top-quark pair production via polarized and unpolarized
photon collisions at the International Linear Collider in the
context of the littlest Higgs model with T-parity. We calculate the
production cross section of the process $\gamma\gamma\rightarrow
t\bar{t}$ and find the effects can be more significant in the $- -$
polarized photon collision mode than in other collision modes, and
the relative correction can be expected to reach about $-1\%$ in the
favorable parameter space.

\end{abstract}
 \maketitle
\section{ Introduction}
\noindent The detailed analysis of the dynamics of top-quark
production and decay is a major objective of experiments at the
Tevatron, the Large Hadron Collider(LHC), and a possible
International Linear Collider(ILC). Top quarks can be produced at
hadron colliders via top-antitop pair \cite{1} and single top
\cite{2} production channels. For top-pair production, the
leading-order (LO) partonic processes are $q\bar{q}\rightarrow
t\bar{t}$, which is dominant at Tevatron energies, and
$gg\rightarrow t\bar{t}$, dominant at LHC energies. At the ILC, one
of the most important reactions will be top-pair production well
above the threshold. Because of the small statistics, the top-quark
properties have not been precisely measured at the Tevatron. Several
tens of millions of top pair signals per year will be produced at
the LHC, this large number of top quarks allow very precise
measurements in the top-quark properties. Compared to the LHC, the
$t\bar{t}$ production cross section is less at the ILC\cite{3}, but
it will be an ideal place for further investigating the top quark
duo to its clean background.

The little Higgs theory constructs the Higgs as a Pseudogoldstone
boson to solve the hierarchy problem of the Standard
Model(SM)\cite{4}. The littlest Higgs (LH) model \cite{5} is an
economical approach to implement this idea, but electroweak
precision tests give it strong constraints \cite{6} so that the
fine-tuning problem in the Higgs potential would be
reintroduced\cite{7}. Then, a discrete symmetry called T-parity is
proposed to tackle this problem\cite{8}, this resulting model is
referred to as the littlest Higgs model with T-parity (LHT).

The LHT model predicts new particles, such as T-odd gauge bosons and
T-odd fermions. In the LHT model, there are interactions between the
SM fermions and the mirror fermions mediated by the new T-odd gauge
bosons or T-odd Goldstone bosons. These interactions can contribute
to the $\gamma t\bar{t}$ coupling and the production cross section
of the process $\gamma\gamma\rightarrow t\bar{t}$. Furthermore, an
additional heavy quark $T^{+}$ and its partner $T^{-}$ can also
contribute to the $\gamma t\bar{t}$ coupling. In this paper, we will
study the polarized and unpolarized $\gamma\gamma$ collisions in the
LHT model at the ILC.

This paper is organized as follows. In Sec.II we give a brief review
of the LHT model. In Sec.III and IV we respectively calculate the
one-loop contributions of the LHT model to the
$\gamma\gamma\rightarrow t\bar{t}$ in polarized and unpolarized
photon-photon collision modes and show some numerical results at the
ILC. Finally, we give our conclusions in Sec.V.

\section{ A brief review of the LHT model}
 \noindent The LHT model \cite{8} is based on an $SU(5)/SO(5)$ non-linear sigma model, where with the
global group $SU(5)$ being spontaneously broken into $SO(5)$ by a
$5\times5$ symmetric tensor at the scale $f\sim\mathcal{O}(TeV)$,
the gauged subgroup $[SU(2)\times U(1)]_{1}\times[SU(2)\times
U(1)]_{2}$ of $SU(5)$ is broken to the diagonal subgroup
$SU(2)_{L}\times U(1)_{Y}$ of $SO(5)$.

From the symmetry breaking, there arise 4 new heavy gauge bosons
$W_{H}^{\pm},Z_{H},A_{H}$ whose masses up to $\mathcal
O(\upsilon^{2}/f^{2})$ are given by
\begin {equation}
M_{W_{H}}=M_{Z_{H}}=gf(1-\frac{\upsilon^{2}}{8f^{2}}),M_{A_{H}}=\frac{g'f}{\sqrt{5}}
(1-\frac{5\upsilon^{2}}{8f^{2}})
\end {equation}
with $g$ and $g'$ being the SM $SU(2)$ and $U(1)$ gauge couplings,
respectively.

A consistent viable implementation of T-parity in the fermion sector
requires the introduction of mirror fermions. For each SM quark, a
copy of mirror quark with T-odd quantum number is added. We denote
up-type and down-type mirror quarks by $u_{H}^{i},d_{H}^{i}$
respectively, where i= 1, 2, 3 are the generation index, whose
masses up to $\mathcal O(\upsilon^{2}/f^{2})$ are given by
\begin{equation}
m_{d_{H}^{i}}=\sqrt{2}\kappa_if, m_{u_{H}^{i}}=
m_{d_{H}^{i}}(1-\frac{\upsilon^2}{8f^2})
\end{equation}
where $\kappa_i$ are the diagonalized Yukawa couplings of the mirror
quarks.

An additional heavy quark $T^{+}$ is introduced to cancel the large
contributions to the Higgs mass from one-loop quadratic divergences.
The implementation of T-parity then requires also a T-odd partner
$T^{-}$, which is an exact singlet under $SU(2)_{1}\times
SU(2)_{2}$. Their masses up to $\mathcal O(\upsilon^{2}/f^{2})$ are
given by
\begin{eqnarray}
m_{T^{+}}&=&\frac{f}{v}\frac{m_{t}}{\sqrt{x_{L}(1-x_{L})}}[1+\frac{v^{2}}{f^{2}}(\frac{1}{3}-x_{L}(1-x_{L}))]\\
m_{T^{-}}&=&\frac{f}{v}\frac{m_{t}}{\sqrt{x_{L}}}[1+\frac{v^{2}}{f^{2}}(\frac{1}{3}-\frac{1}{2}x_{L}(1-x_{L}))]
\end{eqnarray}
where $x_{L}$ is the mixing parameter between the SM top-quark $t$
and the heavy quark $T^{+}$.

In the LHT model, one of the important ingredients of the mirror
sector is the existence of four CKM-like unitary mixing matrices,
two for mirror quarks and two for mirror leptons:
\begin{equation}
V_{Hu},V_{Hd},V_{Hl},V_{H\nu}
\end{equation}
where $V_{Hu}$ and $V_{Hd}$ are for the mirror quarks which are
present in our analysis. They satisfy the relation
$V_{Hu}^{\dag}V_{Hd}=V_{CKM}$. We follow Ref.\cite{9} to
parameterize $V_{Hd}$ with three angles
$\theta^d_{12},\theta^d_{23},\theta^d_{13}$ and three phases
$\delta^d_{12},\delta^d_{23},\delta^d_{13}$
\begin{eqnarray}
V_{Hd}=
\begin{pmatrix}
c^d_{12}c^d_{13}&s^d_{12}c^d_{13}e^{-i\delta^d_{12}}&s^d_{13}e^{-i\delta^d_{13}}\\
-s^d_{12}c^d_{23}e^{i\delta^d_{12}}-c^d_{12}s^d_{23}s^d_{13}e^{i(\delta^d_{13}-\delta^d_{23})}&
c^d_{12}c^d_{23}-s^d_{12}s^d_{23}s^d_{13}e^{i(\delta^d_{13}-\delta^d_{12}-\delta^d_{23})}&
s^d_{23}c^d_{13}e^{-i\delta^d_{23}}\\
s^d_{12}s^d_{23}e^{i(\delta^d_{12}+\delta^d_{23})}-c^d_{12}c^d_{23}s^d_{13}e^{i\delta^d_{13}}&
-c^d_{12}s^d_{23}e^{i\delta^d_{23}}-s^d_{12}c^d_{23}s^d_{13}e^{i(\delta^d_{13}-\delta^d_{12})}&
c^d_{23}c^d_{13}
\end{pmatrix}
\end{eqnarray}

\section{Top quark pair production via $\gamma\gamma$ collision in the LHT model}
\noindent  In the context of the LHT model, the relevant Feynman
diagrams of the one-loop correction to the process
$\gamma\gamma\rightarrow t\bar{t}$ are shown in Fig.1, where the
black dot represents the effective $\gamma t\bar{t}$ vertex which is
shown in Fig.2 and the black diamond represents the fermion
propagator.

In our calculation, the higher order couplings between the scalar
triplet $\Phi$ and top quark and the high order $\mathcal
O(\upsilon^{2}/f^{2})$ terms in the masses of new particles and in
the Feynman rules are neglected. The relevant Feynman rules can be
found in Ref.\cite{13}. We use the 't Hooft-Feynman gauge, so the
masses of the Goldstone bosons and the ghost fields are the same as
their corresponding gauge bosons. The ultraviolet divergences have
been regulated by the dimensional regularization scheme and the
divergences have been canceled according to the on-shell
renormalization scheme.

\begin{figure}[htbp]
\scalebox{0.4}{\epsfig{file=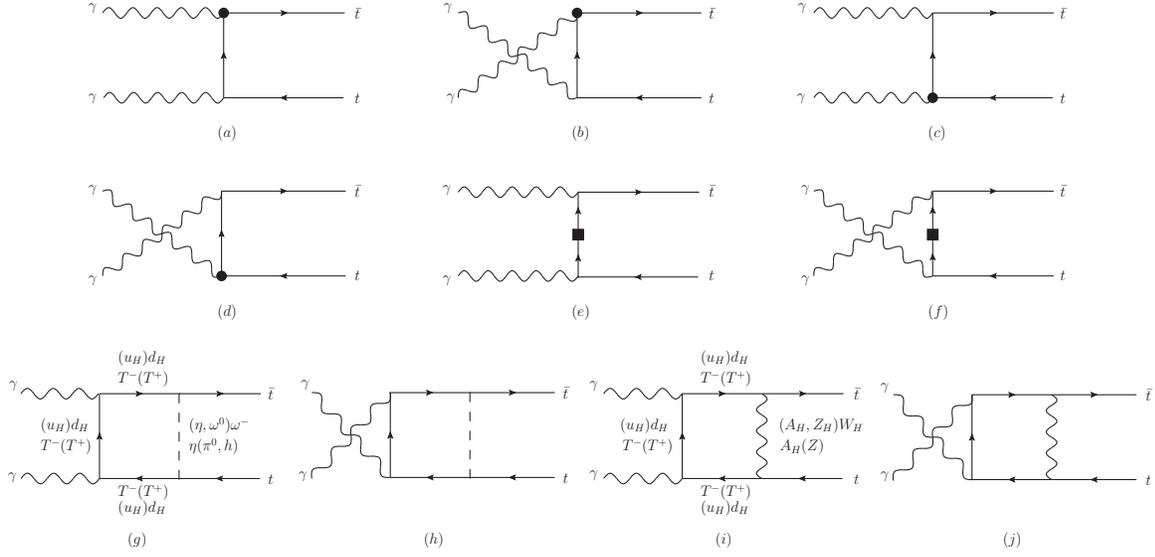}}\caption{Feynman diagrams of
the one-loop correction to the process $\gamma\gamma\rightarrow
t\bar{t}$ in the LHT model.}
\end{figure}
\begin{figure}[htbp]
\scalebox{0.4}{\epsfig{file=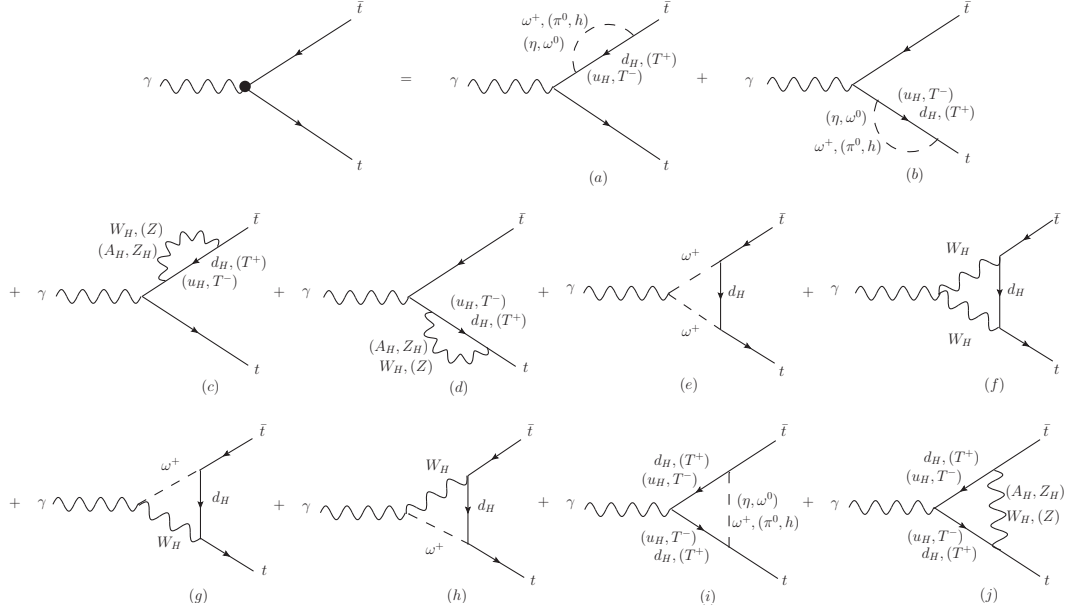}} \caption{The effective $\gamma
t\bar{t}$ vertex diagrams at one-loop level in the LHT model.}
\end{figure}

For the $\gamma \gamma$ collision at the ILC, the photon beams are
generated by the backward Compton scattering of incident electron-
and laser-beams just before the interaction point. The total cross
section $\sigma(s)$ for the top-pair production can be obtained by
folding the elementary cross section $\hat{\sigma}(\hat{s})$ for the
subprocess $\gamma\gamma \rightarrow t\bar{t}$ with the photon
luminosity at the $e^+e^-$ colliders given in Ref.\cite{folding}
\begin{equation}
\sigma(s)=\int_{2m_t/\sqrt{s}}^{x_{\rm max}} {\rm d}z \frac{{\rm
d}L_{\gamma\gamma}}{{\rm d}z}\hat{\sigma}(\hat{s}) \ \ (\gamma\gamma
\rightarrow t\bar{t} \ {\rm at} \ \hat{s}=z^2 s),
\end{equation}
where $\sqrt{s}$ and $\sqrt{\hat{s}}$ are the $e^+e^-$ and
$\gamma\gamma$ center-of-mass energies respectively, and ${\rm
d}L_{\gamma\gamma}/{\rm d}z$ is the photon luminosity, which can be
expressed as
\begin{equation}
 \frac{{\rm
d}L_{\gamma\gamma}}{{\rm d}z}=2z\int_{z^2/x_{\rm max}}^{x_{\rm max}}
\frac{{\rm d}x}{x} F_{\gamma/e}(x)F_{\gamma/e}(z^2/x).
\end{equation}
For unpolarized initial electron and laser beams, the energy
spectrum of the backscattered photon is given by
\begin{eqnarray}
F_{\gamma/e}(x)&=&\frac{1}{D(\xi)} \left ( 1-x+\frac{1}{1-x}-\frac{4
x}{\xi (1-x)}+ \frac{4 x^2}{\xi^2 (1-x)^2} \right ) .
\end{eqnarray}
where $\xi=4E_e E_0/m_e^2$,$m_e$ and $E_e$ are respectively the
incident electron mass and energy, $E_0$ is the initial laser photon
energy. In our numerical calculation, we choose $\xi=4.8$,
$D(\xi)=1.83$ and $x_{max}=0.83$.

The $\gamma\gamma$ collisions have five polarization modes as
follows: $++, --, +-,-+$ and unpolarized collision modes, where the
notation $+$ and $-$ represent the helicities of the two incoming
photons being $\lambda_{1}=1$ and $\lambda_{1}=-1$, respectively.

\section{Numerical results} \noindent
In our numerical calculations, we take the SM parameters as\cite{14}
\begin{eqnarray}
\nonumber G_{F}&=&1.16637\times 10^{-5}GeV^{-2},
~~~S_{W}^{2}=\sin^{2}\theta_{W}=0.231,\\
\alpha_{e}&=&1/128,~M_{Z_{L}}=91.2GeV,~m_{t}=172.4GeV,~m_{h}=120GeV.
\end{eqnarray}

The LHT parameters relevant to our study are the scale $f$, the
mixing parameter $x_{L}$, the mirror quark masses and the parameters
in the matrices $V_{Hu},V_{Hd}$. For the mirror quark masses, we get
$m_{u_{H}^{i}}=m_{d_{H}^{i}}$ at $\mathcal O(\upsilon/f)$ and
further assume
\begin{equation}
m_{u_{H}^{1}}=m_{u_{H}^{2}}=m_{d_{H}^{1}}=m_{d_{H}^{2}}=M_{12},m_{u_{H}^{3}}=m_{d_{H}^{3}}=M_{3}
\end{equation}

From the couplings between the mirror quarks and the heavy gauge
bosons or the heavy Goldstone bosons, we can see the main
contribution comes from the third family couplings. In order to show
the largest correction, for the matrices $V_{Hu},V_{Hd}$, we follow
Ref.\cite{16} to choose the following scenario:
$V_{Hu}=1,V_{Hd}=V_{CKM}$. In this scenario, the contribution of the
LHT model comes entirely from the third family mirror quarks and the
additional heavy quarks $T^{+},T^{-}$.

\begin{figure}[htbp]
\begin{center}
\scalebox{0.7}{\epsfig{file=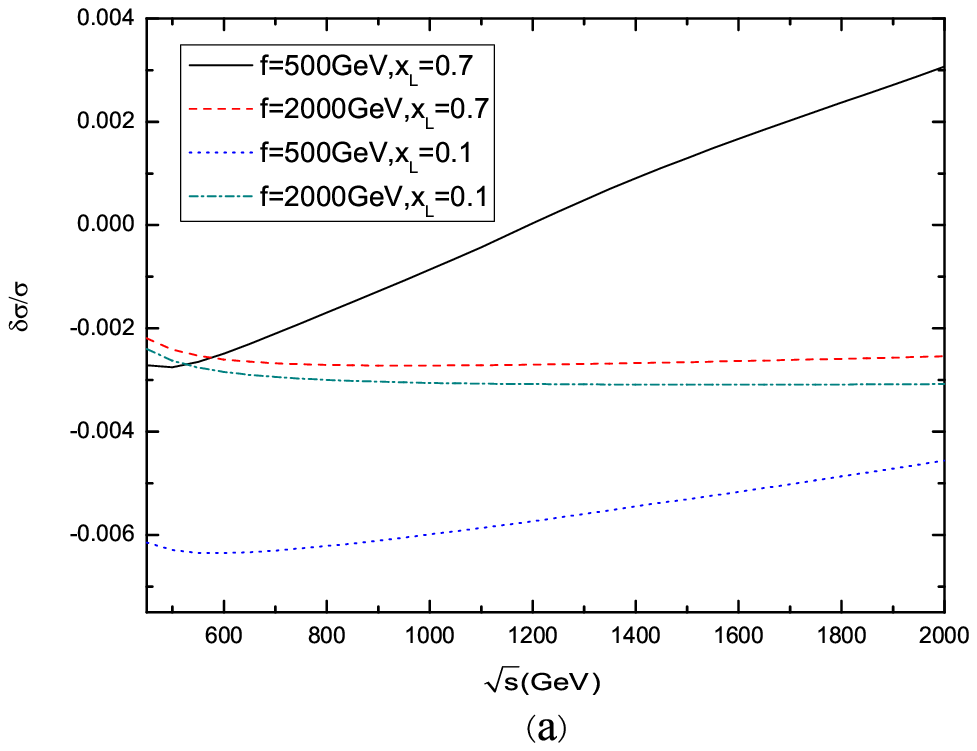}}\vspace{-1cm}
\hspace{-0.5cm} \scalebox{0.7}{\epsfig{file=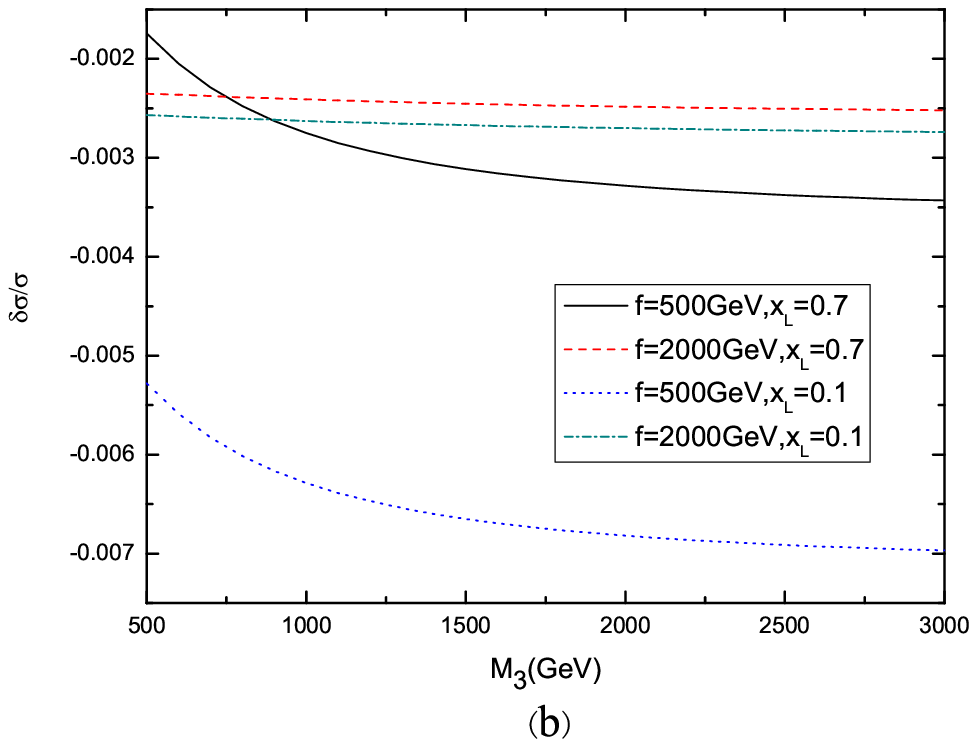}}
\caption{The relative correction of the top-quark pair production
cross section $\delta \sigma/\sigma$ as functions of the
center-of-mass energy $\sqrt{s}$ for $M_{3}=1000GeV$(a) and the
mirror quark mass $M_{3}$ for $\sqrt{s}=500GeV$(b) in unpolarized
photon collision mode, respectively.}
\end{center}
\end{figure}

In Fig.3(a), we discuss the dependance of the relative correction
$\delta \sigma/\sigma$ on the center-of-mass energy $\sqrt{s}$. To
see the influence of the scale $f$ on the $\delta \sigma/\sigma$, we
take $f = 500, 2000GeV$, respectively. We can see $\delta
\sigma/\sigma$ firstly decreases then increases with $\sqrt{s}$. The
higher the scale $f$ is taken, the shallower the curve becomes. To
see the influence of the $x_{L}$ on the $\delta \sigma/\sigma$, we
take $x_{L} = 0.1, 0.7$, respectively. We can see that the larger
the $x_{L}$ is taken, the less the negative relative correction is
generated. Furthermore, the larger the scale $f$ is taken, the less
influence the $x_{L}$ can exert, which means the contributions of
the $T^{+},T^{-}$ are supressed by the high scale $f$. Considering
the constraints of Ref.\cite{constraint}, the maximum of the
relative correction can reach about $-0.65\%$.

In Fig.3(b), we discuss the dependance of $\delta \sigma/\sigma$ on
the third family mirror quark mass $M_{3}$. We can see $\delta
\sigma/\sigma$ is negative and becomes larger with the $M_{3}$
increasing. Same as above, we take $f = 500,2000GeV$ respectively to
see the influence of the scale $f$ on the $\delta \sigma/\sigma$ and
take $x_{L} = 0.1,0.7$ respectively to see the influence of the
$x_{L}$ on the $\delta \sigma/\sigma$. The larger the scale $f$ is
taken, the shallower the curve becomes, which means the
contributions of the $T^{+},T^{-}$ and the third family mirror quark
are all supressed by the high scale $f$. For the same $x_{L}$, the
overall trend is that the higher the scale $f$ is taken, the smaller
the relative correction $\delta \sigma/\sigma$ is generated. For the
same scale $f$, the larger the $x_{L}$ is taken, the less the
relative correction $\delta \sigma/\sigma$ is generated, which means
$\delta \sigma/\sigma$ becomes less with the $M_{T^{+}},M_{T^{-}}$
increasing. Furthermore, we find the contribution of the third
family mirror quark is negative while the contributions of the
$T^{+},T^{-}$ change from negative to positive with the $x_{L}$ from
0.1 to 0.7. For the case $f=500GeV,x_L=0.7$, the contributions of
the $T^{+},T^{-}$ are positive so that they counteract the
contribution of the third family mirror quark very strongly. As a
result, the $\delta \sigma/\sigma$ for $f=500GeV,x_L=0.7$ is smaller
than the case for $f=2000GeV, x_L=0.7$ when $M_{3}<750GeV$. For the
case $f=500GeV,x_L=0.1$, the contributions of the $T^{+},T^{-}$ are
negative so that the contribution of the third family mirror quark
is enhanced obviously. The maximum of the relative correction can
reach about $-0.7\%$.
\begin{figure}[htbp]
\begin{center}
\scalebox{0.7}{\epsfig{file=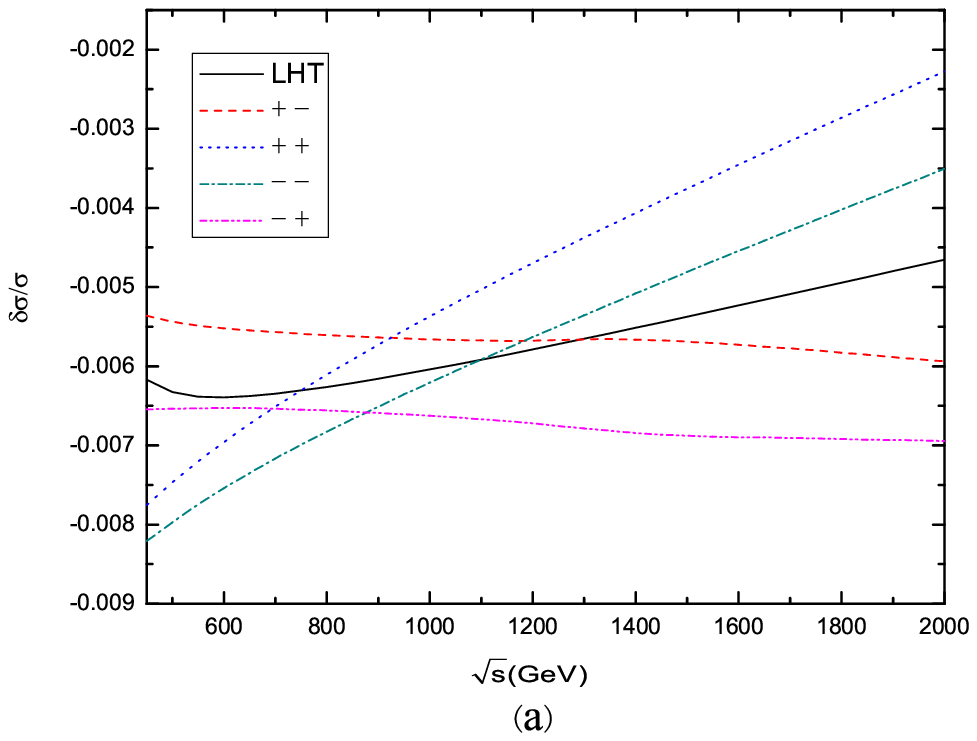}}\vspace{-1cm}
\hspace{-0.5cm} \scalebox{0.7}{\epsfig{file=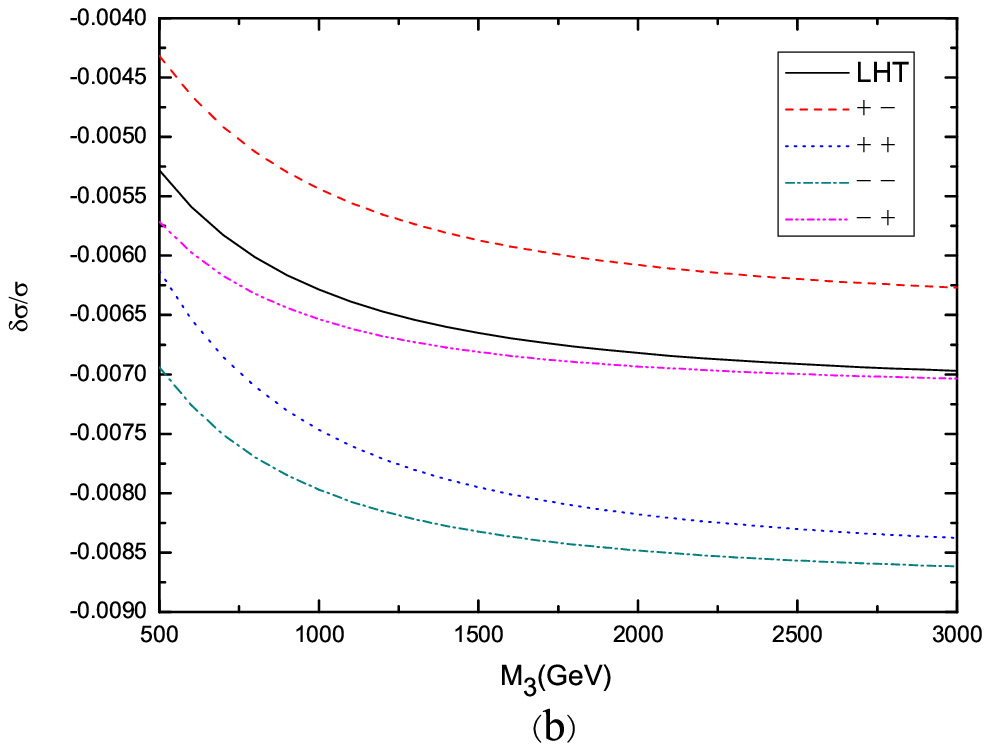}}
\caption{The relative correction of the top-quark pair production
cross section $\delta \sigma/\sigma$ as functions of the
center-of-mass energy $\sqrt{s}$ for $M_{3}=1000GeV, f=500GeV,
x_{L}=0.1$(a) and the mirror quark mass $M_{3}$ for
$\sqrt{s}=500GeV, f=500GeV, x_{L}=0.1$(b), respectively.}
\end{center}
\end{figure}

To see the maximum of the relative correction in the LHT model, we
take $f=500GeV, x_{L}=0.1$ for the process $\gamma\gamma\rightarrow
t\bar{t}$ in polarized photon collision mode. We show the dependance
of the relative correction $\delta \sigma/\sigma$ on the
center-of-mass energy $\sqrt{s}$ and the third family mirror quark
mass $M_{3}$ for the process $\gamma\gamma\rightarrow t\bar{t}$ with
unpolarized and completely $+ -, + +, - -,- +$ polarized photon
beams in Fig4.(a) and Fig4.(b), respectively. From Fig.4(a) and
Fig.4(b) we can see clearly that the relative correction $\delta
\sigma/\sigma$ in $- -$ photon polarization collision mode are
larger than in other photon collision modes. In $- -$ photon
polarization collision mode, the maximum of the relative correction
can be expected to reach about $-1\%$ in the favorable parameter
space.

\section{Conclusions} \noindent In the framework of the LHT model, we studied the
one-loop contributions of the T-odd particles to the top-quark pair
production cross section in unpolarized and polarized photon
collision modes. Because the contributions of the $T^{+},T^{-}$
change from negative to positive with the $x_{L}$ increasing, in
some cases the contribution of the third family mirror quark to the
relative correction $\delta \sigma/\sigma$ was enhanced, and in
other cases the contribution was counteracted. In all collision
modes, we found that the relative correction $\delta \sigma/\sigma$
can be more significant in the $- -$ polarized photon collision mode
than in other collision modes. In the favorable parameter space, the
relative correction can be expected to reach about $-1\%$.

\vspace{1cm}
\textbf{Acknowledgments}\\
We thank Cao Jun-jie for providing the calculation programs and
thank Wu Lei for useful discussions. This work is supported by the
National Natural Science Foundation of China under Grant
Nos.10775039, 11075045, by Specialized Research Fund for the
Doctoral Program of Higher Education under Grant No.20094104110001
and by HASTIT under Grant No.2009HASTIT004.
\begin{center}
\textbf{Appendix: The expression of the renormalization vertex
$\hat{\Gamma}^{\mu}_{\gamma t\bar{t}}$ and the renormalization
propagator $-i\hat{\Sigma}^f(p)$} \cite{17}
\end{center}
(I)Renormalization vertex
\begin{figure}[htbp]
\scalebox{0.4}{\epsfig{file=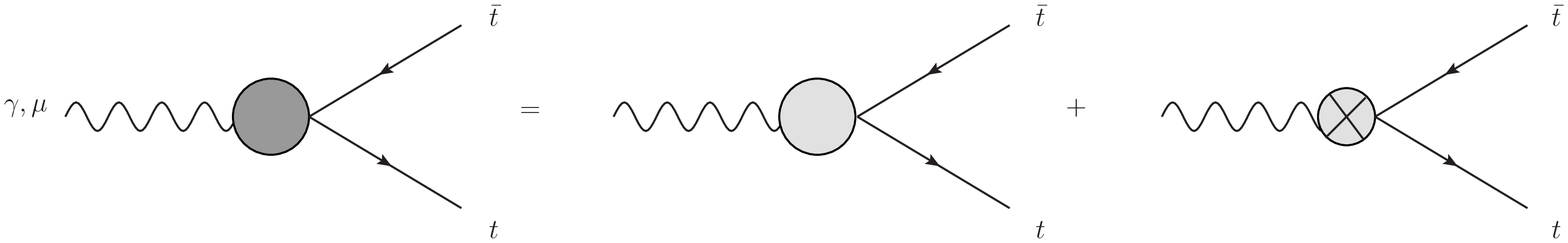}}
\end{figure}
\begin{eqnarray}
\hat{\Gamma}^{\mu}_{\gamma t\bar{t}}&=&\Gamma^{\mu}_{\gamma
t\bar{t}}-ieQ_{t}\gamma^{\mu}(\delta Z_{V}^{t}-\gamma_{5}\delta
Z_{A}^{t}-\frac{S_{W}}{2C_{W}}\delta Z_{ZA})
+ie\gamma^{\mu}(v_{t}-a_{t}\gamma_{5})\frac{1}{2}\delta
Z_{ZA}\nonumber
\end{eqnarray}

where
\begin{eqnarray*}
v_{t}\equiv\frac{I_{t}^{3}-2Q_{t}S_{W}^{2}}{2C_{W}S_{W}},\quad
a_{t}\equiv\frac{I_{t}^{3}}{2C_{W}S_{W}},
\quad I_{t}^{3}=\frac{1}{2},\quad Q_{t}=\frac{2}{3}~~~~~~~~~~~\\
\quad \delta Z_{ZA}=2\frac{\Sigma_{T}^{AZ}(0)}{M_{Z_{L}}^{2}}~~~~~~~~~~~~~~~~~~~~~~~~~~~~~~~~~~~~~~~~~~~~~~~~~~~~~~~~~~~~~~~~~~\\
\delta
Z_{L}^{t}=Re\Sigma_{L}^{t}(m_{t}^{2})+m_{t}^{2}\frac{\partial}{\partial
P_{t}^{2}}Re[\Sigma_{L}^{t}(P_{t}^{2})+\Sigma_{R}^{t}(P_{t}^{2})+2\Sigma_{S}^{t}(P_{t}^{2})]|_{P_{t}^{2}=m_{t}^{2}}\\
\delta
Z_{R}^{t}=Re\Sigma_{R}^{t}(m_{t}^{2})+m_{t}^{2}\frac{\partial}{\partial
P_{t}^{2}}Re[\Sigma_{L}^{t}(P_{t}^{2})+\Sigma_{R}^{t}(P_{t}^{2})+2\Sigma_{S}^{t}(P_{t}^{2})]|_{P_{t}^{2}=m_{t}^{2}}\\
\delta Z_{V}^{t}=\frac{1}{2}(\delta Z_{L}^{t}+\delta
Z_{R}^{t}),\delta Z_{A}^{t}=\frac{1}{2}(\delta Z_{L}^{t}-\delta
Z_{R}^{t})~~~~~~~~~~~~~~~~~~~~~~~~~~~~~~~
\end{eqnarray*}
\begin{eqnarray*}
\hat{\Gamma}^{LHT,\mu}_{\gamma t\bar{t}}&=&\Gamma^{\mu}_{\gamma
t\bar{t}}(\eta)+ \Gamma^{\mu}_{\gamma
t\bar{t}}(\omega^{0})+\Gamma^{\mu}_{\gamma
t\bar{t}}(\omega^{\pm})+\Gamma^{\mu}_{\gamma
t\bar{t}}(\pi^{0})+\Gamma^{\mu}_{\gamma
t\bar{t}}(h)\\&+&\Gamma^{\mu}_{\gamma
t\bar{t}}(A_{H})+\Gamma^{\mu}_{\gamma t\bar{t}}(Z_{H})+
\Gamma^{\mu}_{\gamma t\bar{t}}(W_{H}^{\pm})+\Gamma^{\mu}_{\gamma
t\bar{t}}(Z)+\Gamma^{\mu}_{\gamma
t\bar{t}}(\omega^{\pm},W_{H}^{\pm})\\&+&\delta\Gamma^{\mu}_{\gamma
t\bar{t}}(\eta)+ \delta\Gamma^{\mu}_{\gamma
t\bar{t}}(\omega^{0})+\delta\Gamma^{\mu}_{\gamma
t\bar{t}}(\omega^{\pm})+ \delta\Gamma^{\mu}_{\gamma
t\bar{t}}(\pi^{0})+\delta\Gamma^{\mu}_{\gamma
t\bar{t}}(h)\\&+&\delta\Gamma^{\mu}_{\gamma
t\bar{t}}(A_{H})+\delta\Gamma^{\mu}_{\gamma t\bar{t}}(Z_{H})+
\delta\Gamma^{\mu}_{\gamma
t\bar{t}}(W_{H}^{\pm})+\delta\Gamma^{\mu}_{\gamma t\bar{t}}(Z)
\end{eqnarray*}
(II)Renormalization propagator
\begin{figure}[htbp]
\scalebox{0.4}{\epsfig{file=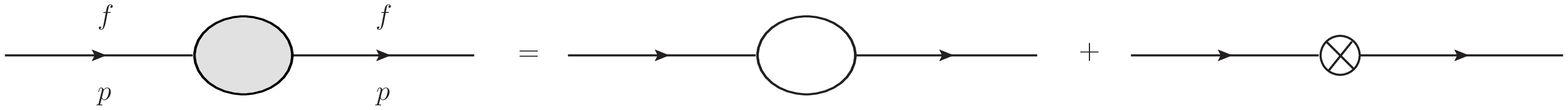}}
\end{figure}
\begin{eqnarray*}
-i\hat{\Sigma}^{f}(p)=-i\Sigma^{f}(p)+(-i\delta\Sigma^{f}(p))
\end{eqnarray*}
where
\begin{eqnarray*}
\Sigma^{f}(p)&=&m_{f}\Sigma^{f}_{S}(p^{2})+\pslash P_{L}\Sigma^{f}_{L}(p^{2})+\pslash P_{R}\Sigma^{f}_{R}(p^{2})\\
\delta\Sigma^{f}(p)&=&\delta m_{f}+m_{f}\frac{1}{2}\delta Z_{L}^{f}+m_{f}\frac{1}{2}\delta Z_{R}^{f}-\pslash P_{L}\delta Z_{L}^{f}-\pslash P_{R}\delta Z_{R}^{f}\\
\delta
m_{f}&=&-m_{f}Re[\Sigma^{f}_{S}(m_{f}^{2})+\frac{1}{2}\Sigma^{f}_{L}(m_{f}^{2})+\frac{1}{2}\Sigma^{f}_{R}(m_{f}^{2})]\\
\delta
Z_{L}^{f}&=&Re\Sigma^{f}_{L}(m_{f}^{2})+m_{f}^{2}\frac{\partial}{\partial
p^{2}}Re[\Sigma^{f}_{L}(p^{2})+\Sigma^{f}_{R}(p^{2})+2\Sigma^{f}_{S}(p^{2})]|_{p^{2}=m_{f}^{2}}\\
\delta
Z_{R}^{f}&=&Re\Sigma^{f}_{R}(m_{f}^{2})+m_{f}^{2}\frac{\partial}{\partial
p^{2}}Re[\Sigma^{f}_{L}(p^{2})+\Sigma^{f}_{R}(p^{2})+2\Sigma^{f}_{S}(p^{2})]|_{p^{2}=m_{f}^{2}}
\end{eqnarray*}


\end{CJK*}

\begin{thebibliography}\\
\bibitem{1}CDF Coll., F. Abe et al., Observation of Top Quark Production in Pbar-P Collisions with the Collider
Detector at Fermilab, Phys. Rev. Lett. 74 (1995) 2626
[hep-ex/9503002]; D0 Coll., S. Abachi et al., Observation of the Top
Quark, Phys. Rev. Lett. 74 (1995) 2632  [hep-ex/9503003].
\bibitem{2}D0 Coll., V.M. Abazov et al., Observation of Single Top-Quark Production, Phys. Rev. Lett. 103 (2009)
092001, arXiv:0903.0850 [hep-ex]; CDF Coll., T. Aaltonen et al.,
Observation of Electroweak Single Top-Quark Production, Phys. Rev.
Lett. 103 (2009) 092002, arXiv:0903.0885 [hep-ex].
\bibitem{3}G. Weiglein et al. [LHC/LC Study Group], Phys. Rept. 426 (2006) 47
 [arXiv:hep-ph/0410364]; J. A. Aguilar-Saavedra et al.
[ECFA/DESY LC Physics Working Group], arXiv:hep-ph/0106315.
\bibitem{4}N. Arkani-Hamed, A. G. Cohen, and H. Georgi, Phys. Lett. B
513 (2001) 232; N. Arkani-Hamed, et al., JHEP 0208 (2002) 020; JHEP
0208 (2002) 021; I. Low, W. Skiba, and D.Smith, Phys. Rev. D 66
(2002) 072001; D. E. Kaplan and M. Schmaltz, JHEP 0310 (2003) 039.
\bibitem{5}N. Arkani-Hamed, A. G. Cohen, E. Katz, and A. E. Nelson,  JHEP
0207 (2002) 034; S.Chang, JHEP 0312 (2003) 057; T. Han, H. E. Logan,
B. McElrath, and L. T. Wang, Phys.Rev. D 67 (2003) 095004; M.
Schmaltz and D. Tucker-smith,  Ann. Rev. Nucl. Part. Sci. 55 (2005)
229.
\bibitem{6}C.Csaki, J.Hubisz, G.D.Kribs, P.Meade, J.Terning, Phys. Rev. D 67 (2003) 115002;  Phys. Rev. D 68 (2003) 035009; J.
L. Hewett,F. J. Petriello, and T. G. Rizzo,  JHEP 0310 (2003) 062;
M. C. Chen and S. Dawson,  Phys.Rev. D 70 (2004) 015003; M. C. Chen,
et al.,  Mod. Phys. Lett. A 21 (2006) 621; W. Kilian, J. Reuter,
 Phys. Rev. D 70 (2004) 015004.
\bibitem{7}G. Marandella, C. Schappacher and A. Strumia,  Phys. Rev. D
72 (2005) 035014.
\bibitem{8}H. C. Cheng and I. Low,  JHEP 0309 (2003) 051;  JHEP 0408 (2004) 061; I. Low,  JHEP 0410 (2004) 067; J. Hubisz and P. Meade,  Phys. Rev. D
71 (2005) 035016.
\bibitem{9}M.Blanke, et al., Phys. Lett. B 646 (2007) 253.
\bibitem{13}M.Blanke, et al., JHEP 0701 (2007) 066.
\bibitem{folding}I. F. Ginzburg et al., Nucl. Instrum. 219 (1984) 5; V. I. Telnov, Nucl. Instrum. Meth. 294 (1990) 72.
\bibitem{14}C. Amsler, et al., Phys. Lett. B 667 (2008) 1.
\bibitem{16}M.Blanke, et al., JHEP 0705 (2007) 013.
\bibitem{constraint}J. Hubisz, P. Meade, A. Noble, and M. Perelstein, JHEP 0601 (2006) 135; Bingfang Yang, Xuelei Wang, and Jinzhong Han,
 Nucl. Phys. B 847 (2011) 1.
\bibitem{17}W.F.L.Hollik, Fortschr. Phys. 38 (1990) 165-260; A.Denner, Fortschr. Phys.41 (1993) 307-420.
\end{thebibliography}
\end{document}